\documentclass[a4paper,12pt]{article}
\pdfoutput=1
\usepackage{graphicx, rotating}
\usepackage{hyperref}
\usepackage{slashed}
\usepackage{ifpdf}
\ifx\pdfoutput\undefined
\usepackage[dvips,bookmarks=false]{hyperref}	% This is for arXiv.org
\else
\usepackage{hyperref}	% This is for pdftex
\fi
\hypersetup{colorlinks,bookmarksopen,bookmarksnumbered,citecolor=verdes,
linkcolor=blus,pdfstartview=FitH,urlcolor=rossos}
\def\hhref#1{\href{http://arxiv.org/abs/#1}{#1}} % in bibliography
\usepackage{amsmath}
\usepackage{slashed}

 \newcommand{\gr}{\ensuremath{\tilde a}}
\newcommand{\gl}{\ensuremath{\tilde g}}
\newcommand{\sq}{\ensuremath{\tilde{q}}}

\newcommand{\beq}{\begin{equation}}
\newcommand{\eeq}{\end{equation}}
\newcommand{\fig}[1]{~\ref{fig:#1}}

\newcount\Mac  \Mac=1  % devo mettere Mac=1 se sto lavorando sul file Mac
\newcommand{\ifMac}[2]{\ifnum\Mac=1 #1 \else #2 \fi}
\def\putps(#1,#2)(#3,#4)#5#6{\ifnum\Mac=1 \put(#1,#2){\special{picture #5}}
\else  \put(#3,#4){\includegraphics{#6}} \fi}

\newcommand{\One}{\hbox{1\kern-.24em I}}

\newcommand{\GeV}{\,{\rm GeV}}

\newcommand{\eV}{\,{\rm eV}}
\newcommand{\NP}{Nucl. Phys.}

\newcommand{\PL}{Phys. Lett.}
\newcommand{\PR}{Phys. Rev.}
\newcommand{\eq}[1]{~{\rm (\ref{eq:#1})}}

\newcommand{\lascia}[1]{}
\makeatletter
\def\art{\@ifnextchar[{\eart}{\oart}}
\def\eart[#1]#2#3#4#5#6{{\rm #2}, {#3 #4} {\rm (#6) #5} [arXiv:{\hhref{#1}}]}
\def\hepart[#1]#2{{\rm #2, arXiv:\hhref{#1}}}
\newcommand{\oart}[5]{{\rm #1}, {#2 #3} {\rm (#5) #4}}

\newcounter{alphaequation}[equation]
\def\thealphaequation{\theequation\hbox to
0.6em{\hfil\alph{alphaequation}\hfil}}
\def\eqnsystem#1{
\def\@eqnnum{{\rm (\thealphaequation)}}
\def\@@eqncr{\let\@tempa\relax \ifcase\@eqcnt \def\@tempa{& & &} \or
  \def\@tempa{& &}\or \def\@tempa{&}\fi\@tempa
  \if@eqnsw\@eqnnum\refstepcounter{alphaequation}\fi
\global\@eqnswtrue\global\@eqcnt=0\cr}
\refstepcounter{equation} \let\@currentlabel\theequation \def\@tempb{#1}
\ifx\@tempb\empty\else\label{#1}\fi
\refstepcounter{alphaequation}
\let\@currentlabel\thealphaequation
\global\@eqnswtrue\global\@eqcnt=0 \tabskip\@centering\let\\=\@eqncr
$$\halign to \displaywidth\bgroup \@eqnsel\hskip\@centering
$\displaystyle\tabskip\z@{##}$&\global\@eqcnt\@ne
\hskip2\arraycolsep\hfil${##}$\hfil& \global\@eqcnt\tw@\hskip2\arraycolsep
$\displaystyle\tabskip\z@{##}$\hfil
\tabskip\@centering&\llap{##}\tabskip\z@\cr}

\def\endeqnsystem{\@@eqncr\egroup$$\global\@ignoretrue} \makeatother

\def\Lag{{\cal L}}

\def\circa#1{\,\raise.3ex\hbox{$#1$\kern-.75em\lower1ex\hbox{$\sim$}}\,}

\usepackage{multicol}
\usepackage{color}
\definecolor{rosso}{cmyk}{0,1,1,0.4}
\definecolor{rossos}{cmyk}{0,1,1,0.55}
\definecolor{rossoc}{cmyk}{0,1,1,0.2}
\definecolor{blu}{cmyk}{1,1,0,0.3}
\definecolor{blus}{cmyk}{1,1,0,0.6}
\definecolor{bluc}{cmyk}{1,1,0,0.1}
\definecolor{verde}{cmyk}{0.92,0,0.59,0.25}
\definecolor{verdec}{cmyk}{0.92,0,0.59,0.15}
\definecolor{verdes}{cmyk}{0.92,0,0.59,0.4}
\definecolor{grigio}{cmyk}{0,0,0,0.07}
\definecolor{rosa}{cmyk}{0,0.1,0.1,0.02}
\definecolor{rosino}{cmyk}{0,0.05,0.05,0.02}
\definecolor{rosas}{cmyk}{0,0.3,0.25,0.05}
\definecolor{celeste}{cmyk}{0.1,0,0,0.02}
\definecolor{giallino}{cmyk}{0,0,0.4,0.02}
\definecolor{rosso}{cmyk}{0,1,1,0.4}
\definecolor{rossos}{cmyk}{0,1,1,0.55}
\definecolor{rossoc}{cmyk}{0,1,1,0.2}
\definecolor{blu}{cmyk}{1,1,0,0.3}
\definecolor{bluc}{cmyk}{1,1,0,0.1}
\definecolor{blucc}{cmyk}{0.7,0.5,0,0}
\definecolor{viola}{cmyk}{0,1,0,0.6}
\definecolor{viola2}{cmyk}{0,1,0.2,0.6}
\definecolor{verde}{cmyk}{0.92,0,0.59,0.25}
\definecolor{verdec}{cmyk}{0.92,0,0.59,0.15}
\definecolor{verdes}{cmyk}{0.92,0,0.59,0.4}
\definecolor{verdino}{cmyk}{0.12,0,0.09,0.05}
\definecolor{giallo}{cmyk}{0,0,1,0}
\definecolor{gialloverde}{cmyk}{0.44,0,0.74,0}

\font\tenrsfs=rsfs10 at 12pt
\font\smallsfs=rsfs10 at 11pt
\font\sevenrsfs=rsfs7
\font\fiversfs=rsfs5
\newfam\rsfsfam
\textfont\rsfsfam=\tenrsfs
\scriptfont\rsfsfam=\sevenrsfs
\scriptscriptfont\rsfsfam=\fiversfs
\def\mathscr#1{{\fam\rsfsfam\relax#1}}
\def\Lag{\mathscr{L}}
\def\Lags{\hbox{\smallsfs L}}

\def\Amp{\mathscr{A}}

\begin{document}
 IFUP-TH/2010-13\hfill CERN-PH-TH/2010-072
\color{black}
\vspace{0.5cm}
\begin{center}
{\Huge\bf\color{rossos}Thermal production\\[3mm] of axino Dark Matter}\\
\bigskip\color{black}\vspace{0.6cm}{
{\large\bf Alessandro Strumia}
} \\[7mm]
{\it CERN, PH-TH, CH-1211, Geneva 23, Switzerland}\\[3mm]
{\it Dipartimento di Fisica dell'Universit{\`a} di Pisa and INFN, Italia}
\end{center}
\bigskip
\centerline{\large\bf\color{blus} Abstract}
\begin{quote}\large
We reconsider thermal production of axinos in the early universe, adding:
a) missed terms in the axino interaction;
b) production via gluon decays kinematically allowed by thermal masses;
c) a precise modeling of reheating.
We find an axino abunance a few times larger than previous computations.

\color{black}
\end{quote}

\section{Introduction}
The strong CP problem can be solved by the Peccei-Quinn symmetry~\cite{PQ}, that manifests at low energy as a light axion $a$
with a decay constant $f\circa{>} 5~10^9\GeV$~\cite{Kim}.
In supersymmetric models the axion $a$ gets extended into an axion supermultiplet
which also contains the scalar saxion $s$ and the fermionic axino $\tilde{a}$~\cite{susyaxion}.
Depending on the model of supersymmetry breaking,
the axino can easily be lighter than all other sparticles, becoming
the stable lightest supersymmetric particle (LSP) and consequently
a Dark Matter candidate~\cite{Wil,Ros,Goto}.
It is thereby interesting to compute its cosmological abundance.
The axino can be produced: i) from decays of the next to lightest sparticle (NLSP),
such that $\Omega_{\tilde{a}} = m_{\tilde{a}}\Omega_{\rm NLSP}/m_{\rm NLSP}$~\cite{Ros}; 
plus ii)
thermally in the early universe when the temperature $T$
was just below the  reheating temperature $T_{\rm RH}$.
We here reconsider the thermal axino abundance, improving on previous
computations~\cite{axinoHTL} in the following ways:
\begin{itemize}
\item[a)] in section~\ref{L} we show that, beyond the well known axino/gluino/gluon interaction,
there is a new axino/gluino/squark/squark interaction, unavoidably demanded by superymmetry,
that contributes at the same order to the usual  $2\to 2$ scatterings that produce axinos;
\item[b)] in section~\ref{T} we show that axinos are also thermally produced
by $1\to 2$ decays kinematically allowed by the gluon thermal mass. The
gluon $\to$ gluino + axino process gives the
dominant contribution in view of the large value $g_3\sim 1$ of the strong coupling constant;
\item[c)] in section~\ref{pheno} we precisely model the reheating process.
\end{itemize}
As a result the axino production rate is significantly enhanced.

\section{Axino couplings}\label{L}
The effective coupling between the axion supermultiplet
$A = (s+ia)/\sqrt{2} + \sqrt{2}\theta \tilde a +\cdots$ and the
strong gauge vectors, described by the gluon $G^a$ and gluino $\tilde{g}^a$
in the  super-multiplet 
%$W_\alpha^a = \tilde{g}_{\alpha}^a + \theta_\beta[\delta_\alpha^\beta D^a -
%i\sigma^{\mu \nu}{}_\alpha{}^\beta G_{\mu \nu}^a ] +\cdots$
$W^a = \tilde{g}^a + \theta[ D^a -
i\sigma^{\mu \nu} G_{\mu \nu}^a ] - i \theta\theta \sigma^\mu D_\mu \bar{\tilde{g}}$ is:
\beq \label{eq:Leff}\Lag_{\rm eff} =-\frac{\sqrt{2}\alpha_3}{8\pi f}  \int d^2\theta\, A W^aW^a + \hbox{h.c.} \eeq
where $\theta$ is the super-space coordinate, $f$ is the axion decay constant, related in a well known way to the parameters of each axion model at hand.
The RGE evolution of $\alpha_3=g_3^2/4\pi$ encodes the RGE renormalization of the full operator. 

By expanding in components and converting to Dirac 4-component notation we get:
%\footnote{Ho cambiato il segno del super-operatore per far 
% tornare i segni dei primi due termini.
%Per i termini con $\lambda$, il fattore di ordine 2 viene tenendo conto che
%Notare che $\int d^2\theta~(\theta\lambda)(\theta\tilde a)  = (\lambda \tilde{a})/2$
%ed \`e in accordo sia con Pradler Steffen che con Covi (che definisce
%$\sigma^{\mu\nu} = i/2 [\gamma^\mu,\gamma^\nu]$).
%Formule varie:
%$$\frac{1}{4}\int d^2\theta WW+h.c.=-\frac{1}{4}FF + \frac{i}{4} F\tilde F + \frac{D^2}{2} - i \bar\lambda \slashed{D}\lambda$$
%$$\sigma^{\nu \mu}{}_\alpha{}^\beta = \frac{1}{4}
%(\sigma^\nu \bar\sigma^\mu-\sigma^\mu\bar\sigma^\nu )_\alpha{}^\beta\qquad
%\bar\sigma^{\nu\mu\dot\alpha}{}_{\dot\beta} = \frac{1}{4}
%(\bar\sigma^\nu \sigma^\mu-\bar\sigma^\mu\sigma^\nu )^{\dot\alpha}{}_{\dot\beta}$$
%Nella notazione di Dirac
%$$
%\gamma^\mu=\left(\begin{matrix} 0 & \sigma^\mu \cr  \bar\sigma^\mu & 0 \cr \end{matrix}\right)
%\qquad
%\gamma_5 = i \gamma^0 \gamma^1 \gamma^2 \gamma^3 = \left(\begin{matrix}-1 &0 \cr
%0 &1 \cr \end{matrix}\right) 
%$$}
%
%\beq \Lag_{\rm eff} =  \frac{\alpha_3}{8 \pi f}\bigg[
%a (G^a_{\mu\nu}\tilde{G}^a_{\mu\nu}+\cdots) +s (G^a_{\mu\nu}G^a_{\mu\nu}+\cdots)
%+ i\,\bar{\tilde a}
%G_{\mu\nu}^a \frac{[\gamma^{\mu },\gamma^{\nu}]}{2}
%\gamma^{5}\tilde{g}^a+2g_3\bar{\tilde{a}} \tilde{g}^a \sum\tilde q^{\ast}T^{a}\tilde q  \bigg]\ .
%\eeq
\begin{eqnarray} \Lag_{\rm eff} &= & \frac{\alpha_3}{8 \pi f}\bigg[
a (G^a_{\mu\nu}\tilde{G}^a_{\mu\nu}+D_\mu(\bar{\tilde{g}}^a \gamma_\mu\gamma_5 \tilde{g}^a)) +
s (G^a_{\mu\nu}G^a_{\mu\nu}-2 D^a D^a+4 \bar{\tilde{g}}^ai\slashed{D}\tilde{g}^a)+  \nonumber \\
&& + i\,\bar{\tilde a}
G_{\mu\nu}^a \frac{[\gamma^{\mu },\gamma^{\nu}]}{2}
\gamma_{5}\tilde{g}^a-2 \bar{\tilde{a}} \tilde{g}^a D^a  \bigg],\qquad 
D^a =-g_3\sum_{\tilde{q}}  \tilde{q}^* T^a \tilde{q} \ .
\end{eqnarray}
The first term is the usual axion coupling to the gluon; the second term is the corresponding saxion coupling
and both are accompanied by couplings to gluinos and to squarks.
%(the $\cdots$ denote terms involving the gluino and the $D$-term); 
The terms in the lower row are the axino couplings; the latter 
term was not considered in the literature~\cite{Wil, axinoHTL,Covi}.
It contributes to the axino production rate at the same order as the well known third term,
as can be seen by inserting the 
explicit value of the strong $D$-term, where the sum runs over all squarks $\tilde{q}$.

\bigskip

In addition to eq.\eq{Leff},
the axion supermultiplet can have extra non-minimal couplings to the electroweak vectors~\cite{Wil,Covi}.  We only consider the presumably dominant strong interaction contribution to the
thermal axino production rate.

\section{Thermal axino production rate}\label{T}

%by demanding
%that the produced gravitinos do not destroy Big Bang Nucleosynthesis,
%see~\cite{Moroi} for a recent precise study.
%This constraint also applies if sparticles are too heavy to be seen at colliders.

\begin{figure}[t]
\begin{center}
$$\includegraphics[width=0.45\textwidth]{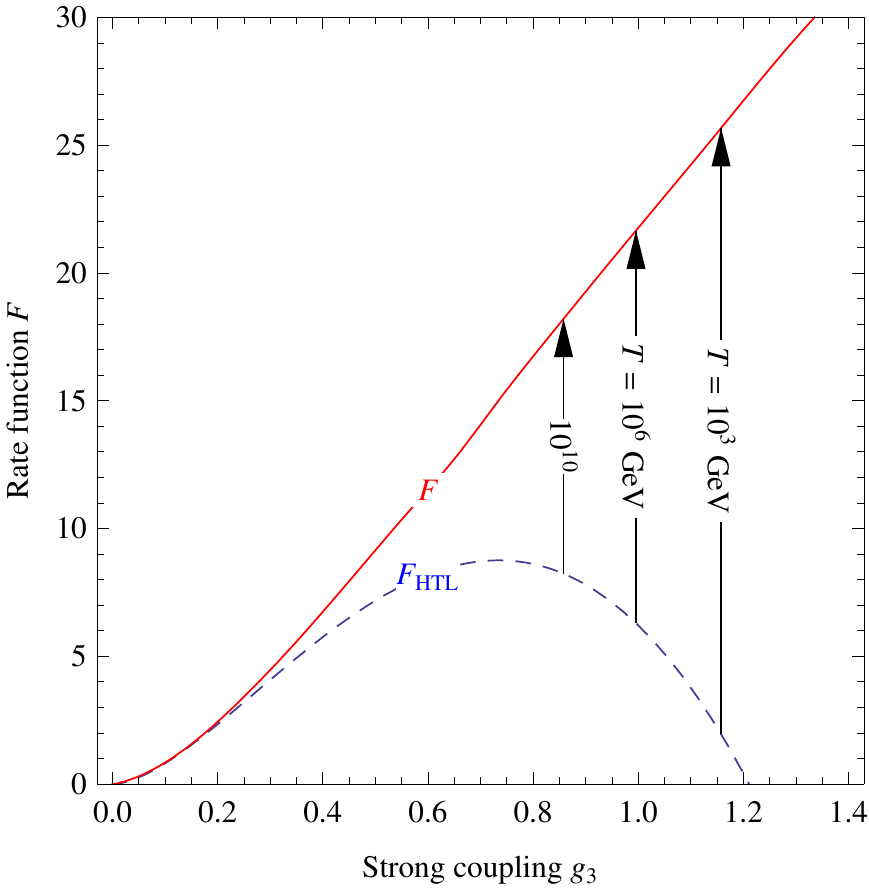}\qquad
\includegraphics[width=0.45\textwidth]{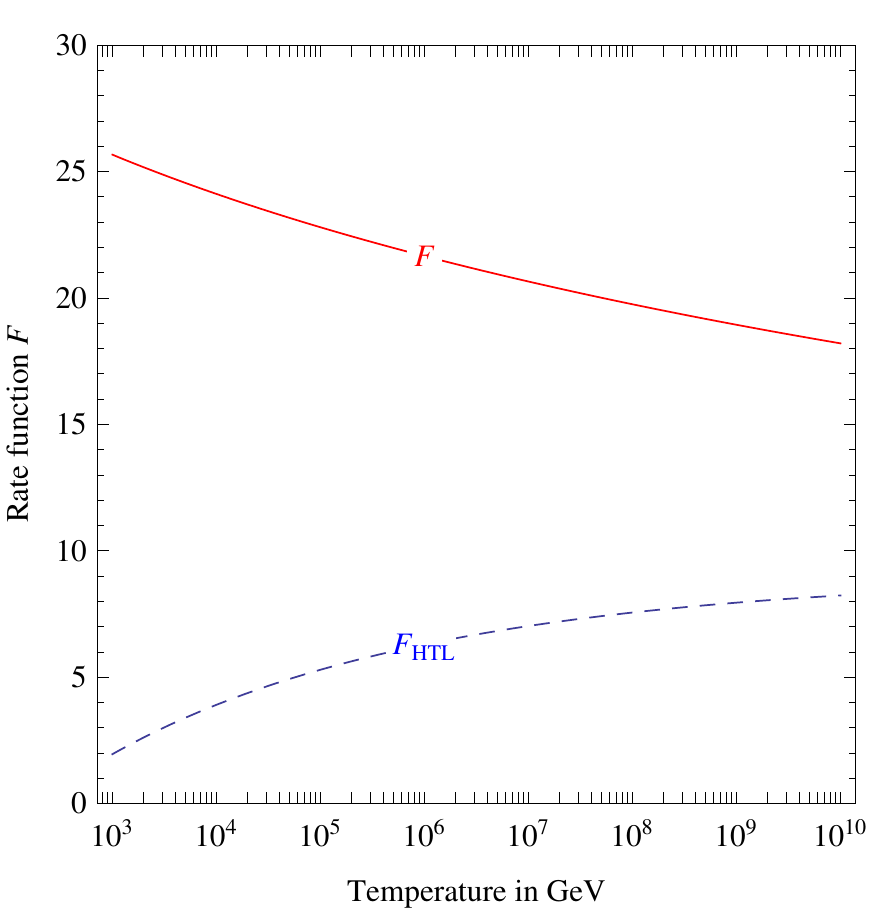}$$
\caption{\label{fig:res}\em Rate function $F$
that describes our result for the axino production rate.
The arrows indicate the MSSM values of $g_3$ at temperatures $T\sim 10^{10-6-3}\GeV$.
The lower dashed curve shows the result $F_{\rm HTL}$
in Hard Thermal Loop approximation
from~\cite{axinoHTL}, which asymptotically
agrees with our result in the limit of small $g_3$, and behaves unphysically for $g_3\sim1 $.
%The upper dashed curve indicates that the Hard Thermal Loop approximation is not accurate for this computation.
%\xxx{FISSARE $g\to 0$}
}
\end{center}
\end{figure}

%Since thermal production of gravitinos already has a vast literature, we start
%presenting the difference with respect to previous computations.
The axino production thermal rate has been computed
in~\cite{axinoHTL} at leading order in the strong gauge
coupling $g_3$. This roughly amounts to compute the $2\to 2$ scatterings listed in table~\ref{tab:diffcs}, 
with thermal effects ignored everywhere but in the propagator of virtual
intermediate gluons: indeed a massless gluon exchanged in the $t$-channel
gives an infinite cross-section because it mediates a long-range
Coulomb-like force; the resulting infra-red logarithmic divergence is cut off
by the thermal mass of the gluon, $m\sim g_3T$, leaving a $\ln T/m$.
The Hard Thermal Loop (HTL~\cite{HTL, leBellac}) approximation ($m\ll T$ i.e.\ $g_3\ll 1$) gives the following result for
the space-time density of scatterings into axinos~\cite{axinoHTL}:
\beq\label{eq:Buch}
 \gamma_{2\to 2} =\frac{T^6 g_3^4}{256\pi^7 f^2} F_{\rm HTL}(g_3),\qquad
 F_{\rm HTL}(g_3) = 32.4\, g_3^2 \ln \frac{1.2}{g_3}\ .\eeq
%\beq\label{eq:Buch}
% \gamma_{\rm scattering} =7.3 \frac{3\zeta(3)T^6}{16\pi^5 \bar M_{\rm Pl}^2}\left(1+\frac{M_3^2}{3 m_{3/2}^2}\right)
%g_3^2 \ln \frac{1.2}{g_3}\eeq
This  production rate unphysically decreases for $g_3\circa{>} 0.7$ becoming
negative for $g_3\circa{>}1.2$~\cite{axinoHTL}.
Fig.\fig{res} illustrates that the physical value, $g_3\approx 0.85$ at $T\sim 10^{10}\GeV$,
lies in the region where the leading-order rate function $F_{\rm HTL}(g_3)$ (dashed line) is unreliable. 
Fig.\fig{res} also illustrates our final result: 
$F_{\rm HTL}$ is replaced by the function $F$ (continuous lines);
they agree at $g_3  \ll 1$ and differ at $g_3\sim 1$.

\medskip

To improve the computation going beyond the leading-order HTL approximation, we notice that
(analogously to what discussed in~\cite{RS} for gravitino thermal production) the new decay process 
\beq\hbox{gluon $\to$ gluino + axino}\label{eq:D}\eeq
first contributes to the axino production rate $\gamma$ at next order in $g_3$.  
Indeed this process is made kinematically allowed by the gluon thermal mass.
Despite being higher order in $g_3$, the decay rate is enhanced by a phase space factor $\pi^2$,
because  a $1\leftrightarrow 2$ process has a phase space larger by this amount than  $2\leftrightarrow 2$ scatterings.
Thereby such decay gives a correction of relative order $(\pi g_3)^2$ to the axino production rate $\gamma$.
Subsequent higher order corrections should be suppressed by the usual $g_3/\pi$ factors.
Our goal is including the enhanced higher order terms, and this finite-temperature computation is
practically feasible  because a decay is a simple enough process.

\begin{table}
\begin{center}%
\begin{tabular}
[c]{c|r@{~$\to$~}l|ccc}
& \multicolumn{2}{c|}{process} & $|\Amp|^{2}_{\mathrm{full}}$ & $|\Amp|^{2}%
_{\mathrm{subtracted}}$ & \\\hline\hline F & $\gl \gl $ & $\gl \gr $
& $-8C{(s^{2}+t^{2}+u^{2})^{2}}/{stu}$ & 0 &
\\\hline
A & $g g $ & $\gl \gr $ & $\phantom{+}4C(s+2t+2t^{2}/s)$ & $-2sC$ &
\\\hline B & $g \gl $ & $g \gr $ & $-4C(t+2s+2s^{2}/t)$ &
$\phantom{+}2tC$ & \\\hline H & $\sq \gl $ & $\sq \gr $ &
$-2C^{\prime}(t+2s+2{s^{2}}/{t})$ & $-tC^{\prime}$ & \\\hline
J & $\sq \bar{\sq} $ & $\gl \gr $ & $\phantom{+}2C^{\prime}(s+2t+2{t^{2}}%
/{s})$ & $\phantom{+}sC^{\prime}$ & \\\hline C & $\sq g $ & $q \gr $
& $\phantom{+}2sC^{\prime}$ & 0 & \\\hline D & $g q $ & $\sq \gr $ &
$-2tC^{\prime}$ & 0 & \\\hline E & $\bar{\sq} q $ & $g \gr $ &
$-2tC^{\prime}$ & 0 & \\\hline G & $q \gl $ & $q \gr $ &
$-4C^{\prime}(s+{s^{2}}/{t})$ & 0 & \\\hline I & $q \bar{q} $ & $\gl
\gr $ & $-4C^{\prime}(t+{t^{2}}/{s})$ & 0 & \\\hline
\end{tabular}
\medskip
\end{center}
\caption{\textit{Squared matrix elements for axino
production in units of $g_3^6/128\pi^4 f^2$
%$........g^{2}_{N}/\bar{M}^{2}_{\mathrm{Pl}} (1+M^{2}_{N}/3m_{3/2}^{2})$, 
summed over all polarizations and gauge
indices.  $g,\tilde{g}, q,\tilde{q}$ denote gluons, gluinos,
quarks, squarks. 
The gauge factors equal
$C=|f^{abc}|^{2}=24$ and
$C^{\prime}=\sum_{{q}}|T_{ij}^{a}|^{2}=48$ after summing over all quarks. }}%
\label{tab:diffcs}%
\end{table}

\medskip

Proceeding along the lines of~\cite{RS},
the axino production rate is precisely defined in terms of the
imaginary part of the thermal axino propagator at one loop~\cite{leBellac},
that we compute using the resummed finite-temperature propagators for gluons and gluinos in the loop.
The resulting expression can be interpreted as the thermal average of the decay process\eq{D},
taking into account that the gluon and gluino thermal masses break Lorentz invariance,
and that actually a continuum of `masses' is present, as precisely described by the
gluon and gluino thermal spectral densities~\cite{leBellac}.

Thermal field theory is just a tool to describe the collective effect of scatterings, and
the decay diagram indeed resums an infinite subset of scatterings, and in particular
some  lowest-order $2\to 2$ scatterings.
Thereby, in order to avoid overcounting,  $2\to 2$ scatterings must be added subtracting the contributions already
included in resummed decay~\cite{RS}, which includes the modulus squared of
single  Feynman $2\to 2$ diagrams.
What remains are interferences between different Feynman $2\to 2$ diagrams~\cite{RS}.
Table~\ref{tab:diffcs} gives explicit values for the total and subtracted
axino
scattering rates. Unlike the total rate, {the subtracted rate is
infra-red convergent}: no $1/t$ factors appear because all divergent
Coloumb-like scatterings are included in the decay diagram.
%In Feynman gauge, rates for the
%processes A and B (the ones that involve two vectors) actually are
%the sum of scatterings involving two vectors (four diagrams,
%computed with the Feynman polarization tensor
%$\sum\epsilon_{\mu}\epsilon_{\nu}^{\ast}=-\eta_{\mu\nu}$) plus
%scatterings containing two ghosts (one diagram, negative
%$|\Amp|^{2}$).
Subtracted rates for processes C, D, E, G, I vanish: a single diagram
contributes, such that no interference terms exist. This is not the
case for scatterings H and J, where the second axino interaction in
eq.\eq{dS} contributes, changing the coefficient of the first term
among parenthesis with respect to the corresponding table~1 of~\cite{axinoHTL},
as well as giving a non-vanishing subtracted contribution.

%In case of scattering F
%the subtracted rate vanishes because proportional to $s+t+u=0$. 
%When deriving the axino production rate, a $1/2!$ factor must be included for the A and F processes that have
%equal initial state particles, and a factor 2 for C, D, G, H that
%can occur with particles and with anti-particles. 

\medskip

Actually there is a stronger connection between the axino and the gravitino thermal production rate.
The latter contains the contribution
of the Goldstino $\chi$ production rate, induced by its coupling 
to the divergence of the MSSM super-current $S^\mu$:
\begin{equation}
\Lags_{\text{int}}=\frac{\bar{\chi}\,\partial_{\mu}S^{\mu}}{\sqrt{2}F}\qquad
\qquad
\partial_{\mu}S^{\mu}=M_3 \bigg[-\frac{i}{4}G_{\mu\nu}^{a}[\gamma^{\mu
},\gamma^{\nu}]\gamma^{5}\tilde{g}^{a}-g_3\tilde{g}^{a}\sum_{\tilde{q}}\tilde{q}^{\ast}T^{a}\tilde{q}\bigg]
\label{eq:dS}
\end{equation}
having including only the contribution from the gluino mass $M_3$.
By comparing with eq.\eq{Leff} we notice that the Goldstino and the axino couple to the same combination of operators.
Even in the Goldstino case, the second operator in $\partial_{\mu}S^{\mu}$ was initially missed and finally noticed by~\cite{LeeWu}.
While the  goldstino mass is predicted in terms of the supersymmetry-breaking $F$ term
as $m_{3/2} = {F}/{\sqrt{3}\bar{M}_{\text{Pl}}}$, the axino mass $m_{\tilde{a}}$ is a model-dependent free parameter.

In conclusion,
after converting $M_3/F \leftrightarrow \sqrt{2}\alpha_3/4\pi f$ the result of~\cite{RS} for the Goldstino thermal production rate
is translated into the axino production rate:
%%%\footnote{\beq\label{eq:res}
%%% \gamma = \gamma_D +\gamma_S^{\rm sub}=\frac{T^6}{2\pi^3} \frac{   M_3^2}{F^2}[ f_3 + 1.29 \cdot \frac{6}{\pi^2}g_3^2]
%%%\eeq}
\beq\label{eq:gamma}
\gamma = \gamma_{1\to 2}+\gamma_{2\to 2}^{\mathrm{sub}}=
\frac{g_3^4 T^6}{256\pi^7 f^2} F(g_3)\qquad
F= f_3 + 1.29 \cdot \frac{6}{\pi^2}g_3^2\eeq
where $f_3$ was numerically computed in~\cite{RS}; the second term is the contribution of $2\to 2$ subtracted scattering rates of table~\ref{tab:diffcs},
and its numerical factor $1.29$ gives the correction coming from the appropriate Fermi-Dirac and Bose-Einstein
distributions with respect to the Boltzmann approximation.

\medskip

Our result for $F$, plotted in fig.\fig{res}, behaves physically for physical values of $g_3\sim 1$.
\bigskip

%As an aside remark, we notice that AdS/CFT techniques should allow to
%compute the axino production rate  in the strong coupling limit, $g_3\to\infty$, 
%although only within  some theory with higher supersymmetries, maybe not unrealistically
%different from supersymmetric QCD. 
%We expect that at strong coupling the axino rate function $F$ has a finite limit, independent of the number of color $N_c$
%up to $1/N_c$ corrections.
%% i.e. \gamma=const. T^6 where const=N^2 up to 1/N corrections

\begin{figure}[t]
\begin{center}
\parbox{0.49\textwidth}{\includegraphics[width=0.45\textwidth]{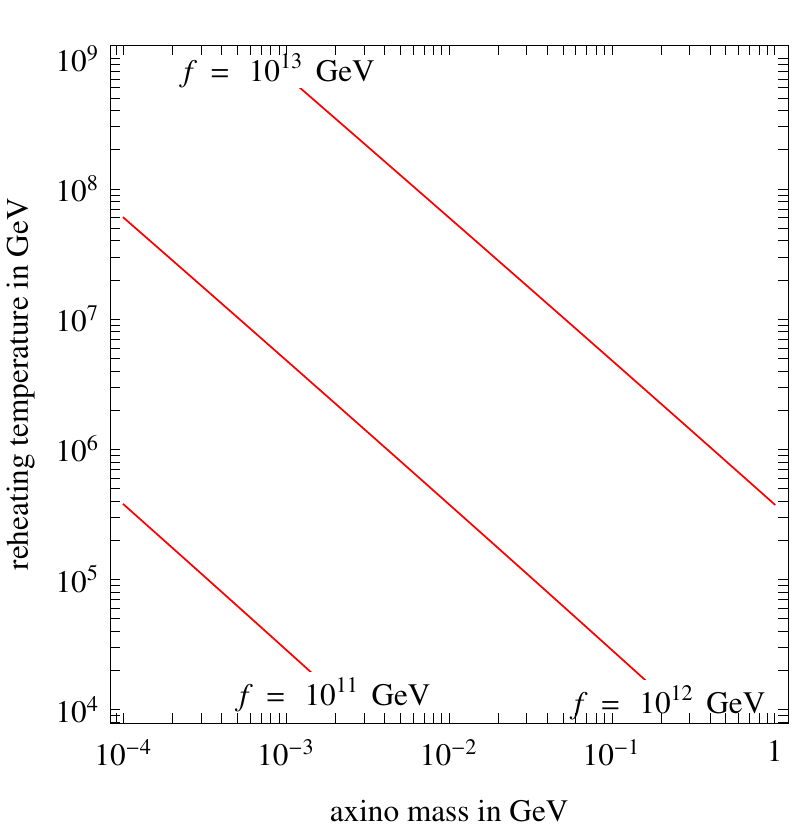}}
\parbox{0.4\textwidth}{
\caption{\label{fig:cos}\em Contour-plot of the value of the axion decay constant $f$ such that
thermally produced axinos are all cold dark matter, $\Omega_{\tilde{a}} = \Omega_{\rm DM}$, 
as function of the axino mass $m_{\tilde{a}}$
and of the reheating temperature $T_{\rm RH}$.
\label{fig:2}
}}
\end{center}
\end{figure}

\section{Thermal axino abundance}\label{pheno}
The axino abundance can now be computed by integrating the relevant cosmological
Boltzmann equation for the axino number density $n_{\tilde{a}}$~\cite{Kh,axinoHTL,RS, Kolb}
\beq \frac{d n_{\tilde{a}}}{dt} + 3 H n_{\tilde{a}} = \gamma\eeq
as well as the equations that describe reheating.
The reheating temperature $T_{\rm RH}$ is approximately defined as the maximal temperature
of the universe (`instantaneous reheating'), 
or more precisely as the temperature at which the inflaton decay rate 
becomes smaller than the Hubble rate $H_R \equiv \sqrt{8\pi \rho_R/3}/M_{\rm Pl}$
(computed including only the energy density $\rho_R$ of the thermal bath), 
such that whatever happened earlier at higher temperatures
got diluted by energy injection from inflaton decay~\cite{Kolb}.
The two different definitions give the following solutions~\cite{RS} for the axino number abundance
normalized to the entropy density $s$:
\beq \label{eq:Yres} \left.\frac{n_{\tilde{a}}}{s}\right|_{T\ll T_{\rm RH}} = \left.
\frac{\gamma}{H_Rs}\right|_{T=T_{\rm RH}} \times\left\{\begin{array}{ll}
0.745 & \hbox{realistic reheating}\\
1 & \hbox{instantaneous reheating}
\end{array}\right.\ .\eeq
Using the numerical factor appropriate for realistic reheating, the present axino mass density  is 
\beq \label{eq:Omega}
\Omega_{\tilde{a}} h^2 =
% 0.00167 \frac{m_{\tilde{a}}}{\GeV}\frac{T_{\rm
%RH}}{10^{10}\GeV} \frac{\gamma|_{T=T_{\rm RH}}}{T_{\rm RH}^6/\bar
%M_{\rm Pl}^2}  = 
1.24  g_3^4F(g_3) \frac{m_{\tilde{a}}}{\GeV}\frac{T_{\rm RH}}{10^{4}\GeV}\bigg(\frac{10^{11}\GeV}{f}\bigg)^2,\qquad
F(g_3) \approx 20 g_3^2 \ln \frac{3}{g_3}
%For such values, $F$ can be numerically approximated as $F \approx 20 g_3^2 \ln 3/g_3$.
 \eeq
with $g_3$ renormalized around $T_{\rm RH}$ and $F(g_3)$ plotted in fig.\fig{res}
(the analytic approximation is appropriate around $g_3\approx 1$).
The axino energy density must be equal or smaller to the DM density $\Omega_{\rm DM} h^2 =0.110\pm 0.006$~\cite{WMAP}, and the resulting phenomenology was studied in~\cite{axinoPheno}.
In fig.\fig{cos} we plot the values of $f$ such that thermally produced axinos have an abundance equal to the observed DM abundance.  We recall that $f\circa{<} 10^{12}\GeV$ in order to avoid a too large axion DM abundance,
unless the initial axion vev is close to the minimum of the axion potential~\cite{axion}.
In eq.\eq{Yres} and\eq{Omega} we assumed that $n_{\tilde{a}} \ll n^{\rm eq}_{\tilde{a}}\approx 1.8~10^{-3}s$, otherwise
the axino reaches thermal equilibrium, giving $\Omega_{\tilde{a}}\gg \Omega_{\rm DM}$ for $m_{\tilde{a}}\gg 10\eV$.

\section{Conclusions}\label{conclusions}

In section~\ref{L} we derived the axino coupling to the strong sector,
finding that it is described by two terms:
one is the well known $\tilde{a} \tilde{g} G$ coupling, the other $\tilde{a}\tilde{g}\tilde{q}^*\tilde{q}$
term was missed in previous studies, although they contribute at the same order to the thermal axino production rate.
 This makes the axino interaction fully analogous to the Goldstino interaction, such
 that we could infer the axino thermal production rate from the Goldstino rate, computed in~\cite{RS}.
Such result goes beyond the leading order computations~\cite{axinoHTL}
 based on the Hard Thermal Loop (HTL) approximation $g_3\ll 1$,
 which gives unphysical results at the physical value of $g_3\sim 1$.
 As a consequence we find an enhancement, plotted in fig.\fig{res},
 by a factor of 6 (3) at $T_{\rm RH}= 10^4~(10^7)\GeV$.
The function $F$ determines the axino rate as described by eq.s\eq{gamma} and\eq{Omega}.

\small

\paragraph{Acknowledgements}
We thank L. Covi, V.S. Rychkov, C. Scrucca and G. Villadoro.

\bigskip

\footnotesize

\begin{multicols}{2}
\end{multicols}
\end{document}